\documentstyle[epsfig]{mn}

\newcommand{\Msolar}{\mbox{\,$\rm M_{\odot}$}}        

\newcommand{\ang}{\mbox{$\rm \AA$}}

\title[The black hole -- bulge mass relation]
{On the black hole -- bulge mass relation in active and inactive galaxies}
\author[R.J. McLure \& J.S. Dunlop]
{R.J. McLure$^1$ \& J.S. Dunlop$^2$\\
 $^{1}$Nuclear and Astrophysics Laboratory, University of Oxford,
Keble Road, Oxford, OX1 3RH\\
$^{2}$Institute for Astronomy, University of Edinburgh, Royal
Observatory, Edinburgh EH9 3HJ.\\}
\date{Accepted for publication in MNRAS}
\begin{document}
\maketitle
\begin{abstract}
New black-hole mass estimates are presented for a
sample of 72 AGN covering three decades in optical luminosity. Using a
sub-sample of Seyfert galaxies, which have black-hole mass estimates from
both reverberation mapping and stellar velocity dispersions, we
investigate the geometry of the AGN broad-line region (BLR). It is
demonstrated that a model in which the orbits of the 
line-emitting material have a flattened geometry is favoured over
randomly orientated orbits. Using this model we investigate 
the $M_{bh}-$~$L_{bulge}$ relation for a combined 90-object sample, 
consisting of the AGN plus a sample of 18 nearby inactive elliptical galaxies 
with dynamical black-hole mass measurements. It is found that, for all
reasonable mass-to-light ratios, the $M_{bh}-L_{bulge}$ relation 
is equivalent to a linear scaling between bulge and black-hole
mass. The best-fitting normalization of the $M_{bh}-M_{bulge}$
relation is found to be $M_{bh}=0.0012M_{bulge}$, in agreement with 
recent black-hole mass studies based on stellar velocity dispersions. 
Furthermore, the scatter around the $M_{bh}-L_{bulge}$
relation for the full sample is found to be significantly smaller than
has been previously reported ($\Delta \log M_{bh}=0.39$ dex). Finally, 
using the nearby inactive elliptical galaxy sample alone, it is shown
that the scatter in the $M_{bh}-L_{bulge}$ relation is only $0.33$ dex,
comparable to that of the $M_{bh}-\sigma$ relation. These results 
indicate that reliable black-hole mass estimates can be obtained for
high redshift galaxies.
\end{abstract}

\section{Introduction}
\label{intro}
The technique of using the broad H$\beta$ emission line to estimate
the central black-hole masses of active galactic nuclei (AGN) has
recently been employed widely
in the literature (eg. Lacy et al. 2001, Laor 2001a, Wandel
1999). Given the ease with which the nuclear spectra of AGN can 
be obtained, a proper calibration of line-width black-hole mass 
estimates has the potential to allow the study of the evolution and 
demographics of active supermassive black holes over a wide range in 
redshift. 

In a previous paper (McLure \& Dunlop 2001, hereafter MD01) we
reported H$\beta$ black-hole mass estimates for a 45-object sample
consisting of both Seyfert galaxies and powerful ($M_{V}<-23$)
quasars. By combining the black-hole mass estimates with host-galaxy 
bulge luminosities derived from full two-dimensional disc/bulge 
decompositions, MD01 found 
that quasars and Seyfert galaxies follow the same 
$M_{bh}-L_{bulge}$ relation, a result recently confirmed by Laor
(2001a). Furthermore, by adopting a simple inclination correction
factor for the H$\beta$ line widths, MD01 found that the 
mean M$_{bh}/$M$_{bulge}$ ratio in AGN
is a factor of $2-4$ lower than had been determined by
Magorrian et al. (1998) for nearby galaxies, in reasonable agreement
with the results from recent stellar velocity dispersion studies
(eg. Merritt \& Ferrarese 2000a). 

Despite the recent attention which has been focussed on determining
the black-hole masses of both active and inactive galaxies, several
important problems remain to be resolved, all of which are
potentially soluble.

Firstly, at present the usefulness of the $M_{bh}-L_{bulge}$ relation as a
black-hole mass estimator for both active and inactive galaxies is 
severely limited due to its large scatter ($\simeq0.5$ dex). In our
previous study (MD01) we demonstrated that much of this scatter could
be due to the difficulty of accurately determining the bulge
luminosities of late-type galaxies, even at $z<0.1$. Although
the correlation between black-hole mass and stellar velocity
dispersion for nearby inactive galaxies displays a much smaller
scatter ($\simeq0.3$ dex, Merritt \& Ferrarese
2000b), it is still clear that a $M_{bh}-L_{bulge}$ correlation with
reduced scatter would be highly desirable, given the extreme difficulty in
obtaining stellar velocity dispersions for high redshift galaxies. 
In this paper we investigate the
possibility that the intrinsic scatter in the $M_{bh}-L_{bulge}$
relation is substantially lower than previously reported by studying a
sample of nearby inactive galaxies for which $L_{bulge}$ has been
determined to high accuracy.

A further complication which arises when studying the $M_{bh}-L_{bulge}$
relation for powerful active galaxies, is the uncertainty in how best to
calibrate the virial black-hole mass estimates produced by using the broad
H${\beta}$ line-width to trace the central gravitational
potential. Indeed, uncertainty about the exact geometry of 
the broad-line region (BLR) in AGN has the potential to 
produce large systematic errors in line-width based black-hole mass
estimates (see Krolik 2000 for a discussion). In this paper we
directly address this issue by adopting a flattened disc-like BLR
geometry, which we demonstrate to be fully consistent with presently
available data. 

Thirdly, it
is currently unclear whether the $M_{bh}-M_{bulge}$ relation is
linear over a large baseline in bulge mass. Although our previous
study (MD01) found no evidence for non-linearity in the
$M_{bh}-M_{bulge}$ relation, the recent study of Laor (2001a) came to
the opposite conclusion, finding that
$M_{bh} \propto M_{bulge}^{1.54}$. As well as being of
intrinsic interest, the question of the linearity of the 
$M_{bh}-M_{bulge}$ relation can also be used to constrain current 
models of combined black hole/bulge formation (eg. Kauffmann \& Haehnelt 2000).

Fourthly, the results of our HST host-galaxy study (McLure et al. 1999,
Dunlop et al. 2001) together with our previous H$\beta$ study
(MD01), and the H$\beta$ study of Laor (2001b), point to an
apparent division between the black-hole masses of
optically selected radio-loud and radio-quiet quasars at 
$\simeq 10^{9}\Msolar$. However, recent results from Lacy et
al. (2001) and Dunlop et al. (2001) both indicate that this apparent
threshold is not of fundamental physical importance. In this paper we 
use a large sample of 72 AGN, together with a
sample of 20 nearby inactive galaxies with dynamic black-hole mass
estimates, to systematically re-address these questions. 

The structure of this paper is as follows. In Section \ref{sample} we
 describe the construction of the AGN and inactive galaxy samples. 
In Section \ref{model} we briefly review the technique of estimating
 black-hole masses from emission-line widths, before proceeding to 
describe our adopted 
disc BLR model, and then to demonstrate that it is
fully consistent with recent velocity dispersion measurements for a
subset of our Seyfert galaxy sample. In Section \ref{main} the
 normalization and linearity of the resulting
$M_{bh}-L_{bulge}$ relation for the AGN and inactive galaxy samples
is investigated and compared with literature results. In this section we
 also compare the $M_{bh}-L_{bulge}$ and $M_{bh}-\sigma$ relations for the
inactive galaxy sample, and demonstrate that the scatter in both is
virtually identical. In Section \ref{leddsec} we 
investigate the correlation between black-hole mass and the 
ratio of bolometric luminosity to the Eddington limit for the AGN
 sample. In Section \ref{dichotomy} we explore the
 implications of the new black-hole mass estimates for the 
radio loudness dichotomy, before
presenting our collusions in Section \ref{conclusions}. 
To provide consistency with our 
previous study, all cosmological calculations in this paper assume
$H_{0}=50$ kms$^{-1}$Mpc$^{-1}$, $\Omega=1.0$, $\Lambda=0$.
\section{The Sample}
\label{sample}
\begin{table}
\begin{center}
\begin{tabular}{lcll}
\hline
Sample&N&$L_{bulge}$&H$\beta$ FWHM\\
\hline
QSO&30&MD01&MD01\\
QSO&11&H97&F01\\
QSO&\phantom{0}8&MPD01&BG92\\
QSO&\phantom{0}4&P01&BG92\\
Sy1&15&MD01&WPM99\\
Sy1&\phantom{0}4&V00&L01\\
Inactive&18&F97&-\\
\hline
\end{tabular}
\caption{Details of the data drawn from the literature. Column 2 lists the 
number of objects in each sub-sample. Column 3 lists the principle 
references for the bulge luminosity data. Column 4 lists the references 
for the H$\beta$ line-width data. The reference codes are as 
follows: MD01 (McLure \& Dunlop
2001), H97 (Hooper et al. 1997), F01 (Forster et al. 2001), MPD01
(McLure, Percival \& Dunlop 2001), BG92 (Boroson \& Green 1992), P01
(Percival et al. 2001), WPM99 (Wandel, Peterson \& Malkan 1999), V00 (Virani 
et al. 2000), L01 (Laor 2001a), F97 (Faber et al. 1997).}
\label{datatab}
\end{center}
\end{table}
The full sample analysed in this paper consists of 90 objects comprising 
three sub-samples of 53 quasars, 19 Seyfert 1 galaxies and 20
inactive nearby galaxies. This sample is a combination of the 
45 objects from our previous study (MD01), together with 
47 additional objects drawn from various literature sources. The principle 
objective behind the construction of the sample was to allow the study of
 the bulge: black-hole mass relationship over the widest possible 
dynamic range in both nuclear luminosity and central black-hole mass. Due 
to the fact that the
sample is drawn from several sources, it was necessary to transform 
all the bulge luminosity data into the same filter. As in MD01, the 
chosen filter was the standard $R$-Cousins, which was preferred because
measured $R-$band bulge luminosities were available for 65/72 of the AGN
sample. The principle references for the bulge luminosity and line-width 
data are provided in Table \ref{datatab}, with specific details 
relating to each sub-sample provided below. 
\subsection{The quasar sample}
The 53 objects in the quasar sample cover the redshift range
$0.1<z<0.5$, and have absolute magnitudes ranging from $M_{R}\sim-22.5$ to
$M_{R}\sim-28.0$. Consequently, the lowest luminosity quasars in the
sample cover the overlap region around the Seyfert/quasar divide,
while the highest luminosity objects constitute some of the most
powerful quasars available at $z\leq0.5$. Thirty of the objects
are drawn from the quasar sample analysed in MD01, which have
 accurate $R-$band bulge luminosities from the host-galaxy 
study of Dunlop et al. (2001), and
have line-width and optical continuum measurements from either our own
recent observations (MD01), or the study of Boroson \& Green
(1992). The remaining 23 quasars in the new sample are drawn from
three additional sources. Eleven further objects are taken from the
HST host-galaxy study of LBQS quasars by Hooper et al (1997). The
bulge luminosities are converted from the $R_{J}$ magnitudes published by
Hooper et al. assuming $R_{C}-R_{J}=0.1$ (Fukugita et al. 1995), with 
the line-width and continuum measurements taken
from Forster et al. (2001). Data for four objects is taken from the
host-galaxy study of high-luminosity radio-quiet quasars by Percival
et al. (2001). These quasars have $K-$band bulge luminosities based on
sub-arcsecond tip/tilt imaging obtained at UKIRT, and optical spectra
taken from Boroson \& Green (1992). The final eight objects are taken
from a new study (McLure, Percival \& Dunlop 2001, in prep) of the
host-galaxies of a radio-loud companion sample to that studied by 
Percival et al. The bulge luminosities in this study are taken from a
combination of new $K-$band UKIRT observations and archive HST data. The
optical spectra for these objects are also taken from Boroson \& Green
(1992). The bulge luminosities for those objects with $K-$band data
were converted to the $R-$band assuming a rest-frame colour of 
$R-K=2.5$ (Dunlop et al. 2001).
\subsection{The Seyfert galaxy sample}
The Seyfert galaxy sample features the 15 objects analysed in MD01,
with the addition of four objects from the Seyfert galaxy study 
of Virani et al. (2000) for which line-width data was
available. In addition to the four new objects, the $R_{C}$ disc/bulge 
decompositions provided by Virani et al. have also been adopted for three 
objects from the MD01 sample (NGC 4051, NGC 4151, and NGC 3227). The
bulge luminosities for these objects from Virani et al. replace those
originally used by MD01, which were taken from Baggett, Baggett \&
Anderson (1998), and are preferred here because the data were obtained
using the $R_{C}$ filter. The bulge luminosities
for the other 12 objects remain the same as in MD01.
\subsection{The inactive galaxy sample}
The normal galaxy sample consists of eighteen objects drawn from the list of
37 nearby inactive galaxies with dynamical black-hole mass 
measurements published by Kormendy \& Gebhardt (2001). Due to
the difficulty in determining accurate bulge luminosities in late-type
galaxies (see MD01 for a discussion) it was decided to exclude those
galaxies in the Kormendy \& Gebhardt list which were not of E-type
morphology (including lenticulars). This decision was taken to 
investigate to what extent the
scatter in the $M_{bh}-L_{bulge}$ relation could be reduced if the
uncertainties in the bulge luminosities were minimized (see Section
\ref{main}). The Kormendy \& Gebhardt list consists of 20 E-type
galaxies, which was reduced to the final 18-object sample
by the exclusion of NGC 4486B and NGC 5845. These two objects were
excluded because the errors on their black-hole mass measurements are
an order of magnitude larger than for the rest of the sample (we note 
that NGC 4486B was also excluded as an outlier by Merritt \& Ferrarese
(2000b) in their study of black-hole demographics). The $B-$band 
bulge luminosities for 9 of the 18 objects were taken from 
Faber et al. (1997), with $V-$band data for a further 9 being taken 
from Merritt \& Ferrarese (2000b). To convert the $B-$band and $V-$band 
magnitudes to the $R-$band, standard bulge colours of 
$B-R=1.57$ and $V-R=0.61$ were assumed (Fukugita et al. 1995).
All eighteen objects in the sample have published velocity dispersion
measurements, allowing a direct comparison of the 
$M_{bh}-L_{bulge}$ and $M_{bh}-\sigma$ relations (see Section \ref{main}).  

\section{The virial black-hole mass estimate}
\label{model}
The technique of using H$\beta$ line-widths to trace the 
gravitational potential of the central 
black holes which power AGN has been used extensively in recent years. 
Detailed discussions are provided by Wandel, Peterson \&\ Malkan (1999) 
and Krolik (2001). The basic premise is that the dominant mechanism
responsible for the width of the broad emission lines is the 
gravitational potential of the central black hole, and that the 
line-widths reflect the keplerian velocities of the 
line-emitting material (see Peterson \& Wandel (2000) for supporting 
evidence). If this assumption is valid then the so-called virial 
mass estimate for the central black hole is given by:
\begin{equation}
M_{bh}= R_{BLR} V^{2} G^{-1} 
\label{eqn1}
\end{equation}
\noindent
where $R_{BLR}$ is the radius of the BLR and $V$ is the velocity
of the line-emitting material.  As in MD01, the method adopted in this
paper for estimating the BLR radius is the correlation 
between $R_{BLR}$ and monochromatic luminosity at $5100\ang$ 
found by Kaspi et al. (2000), from a combination of the reverberation 
mapping results for 17 PG quasars and 17 Seyfert galaxies 
(15 of which are in the sample studied here). Incorporating this into 
Eqn \ref{eqn1} yields the following formula for the 
black-hole mass:
\begin{equation}
M_{bh}=3.64\times f^{2} \times \left( 
\frac{\lambda L_{\lambda}(5100\ang)}{10^{37}W}\right)^{0.7} 
\times (FWHM)^{2}
\label{eqn2}
\end{equation}
\noindent
where $M_{bh}$ is the black-hole mass in solar units, FWHM is the
full-width half maximum of the H$\beta$ line in kms$^{-1}$, and $f$ is a 
inclination factor which links the observed H$\beta$
FWHM to the intrinsic velocity of the line-emitting material. It can
immediately be seen from Eqn \ref{eqn2} that the inclination factor $f$ 
can have a potentially large effect on the derived black-hole 
masses. Although evidence exists that radio-loud AGN display 
inclination dependent FWHMs consistent with a disc-like BLR (see
below), in the radio-quiet population
the geometry of the BLR is arguably undetermined. Consequently, 
the standard assumption made in the literature is
that the velocities of the line-emitting material are randomly
orientated, which leads to $f=\frac{\sqrt{3}}{2}$ (eg. Wandel 1999). 
However, as in our previous study (MD01), in this paper we 
adopt a specific model for the geometry of the broad-line region,
which, as discussed below, is more consistent with the available data.
\subsection{The disc model of the broad-line region}
\label{discsec}
\begin{figure*}
\centerline{\epsfig{file=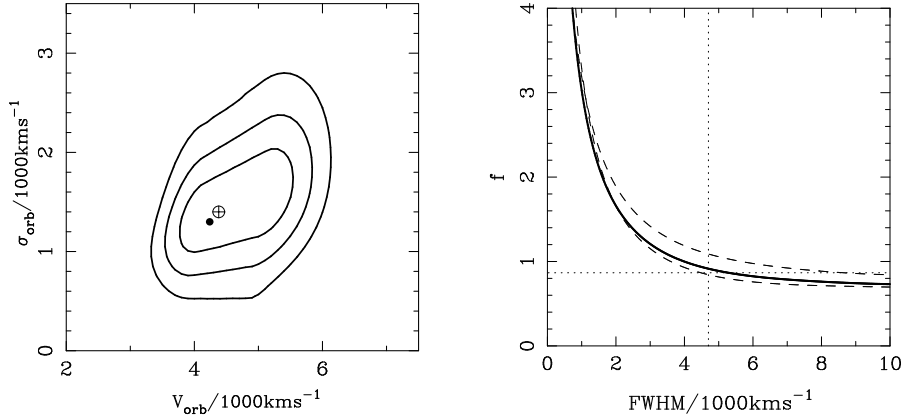,width=12cm,angle=0,clip=}}
\caption{The left-hand panel shows the location (cross) of the 
best-fitting values of $V_{orb}$ and $\sigma_{orb}$, together with 
the $1, 2\,\& \, 3\sigma$ joint confidence levels 
defined by $\Delta \chi^{2}=2.3, 6.2 \,\&\, 11.8$. Also shown 
(filled circle) is the location of the
best-fitting parameters obtained by minimizing the Kolmogorov-Smirnov
statistic between the model and data cumulative FWHM distributions.
The right-hand panel shows the inclination correction factor ($f$) derived from
the best-fitting parameters (solid line), together with the $\pm 1\sigma$
models (dashed lines). Also shown is the effective inclination
factor resulting from the assumption of random broad-line velocities
(horizontal dotted line) and the mean observed FWHM of our sample
(vertical dotted line).}
\label{discmodel}
\end{figure*}
Following MD01 we make the assumption in this paper that the 
broad-line emitting material has a flattened disc-like geometry, and 
consequently that the observed H$\beta$ FWHM depends on the orientation
of the disc rotation axis relative to the line of sight.
Substantial evidence exists in the literature that the FWHM of 
broad emission lines are inclination dependent in radio-loud AGN (eg. 
Wills \& Browne 1986, Brotherton 1996, Corbin 1997, 
Vestergaard, Wilkes \& Barthel 2000). Such strong 
observational evidence does not exist for radio-quiet AGN, largely because 
the absence of radio jets removes one of the main methods with which
to constrain the AGN orientation axis. However, the extension of 
a disc BLR model to the 
radio-quiet AGN in our sample can be justified on several grounds. Firstly, 
under the unified scheme (eg. Urry \& Padovani 1995) it is expected 
that (to first order) the central engines, and presumably BLR, of 
radio-loud and radio-quiet AGN are essentially identical. Secondly, 
as will be discussed further below, 
previous studies have demonstrated that it is possible to fit the broad 
emission-line spectra of both radio-loud and radio-quiet AGN using a 
BLR model with a flattened geometry. Thirdly, as will be seen in 
Section \ref{geometry}, for those 
radio-quiet objects in our sample where it is possible to test the geometry 
of the BLR, it appears that the data are most 
consistent with a flattened geometry.

In order to determine the parameters of the chosen disc model we
simply require that the model can reproduce the 
distribution of the observed H$\beta$ FWHMs. In MD01, the cumulative FWHM
distribution of a smaller 45-object sample was fitted using a 
two-population model, where both sub-populations had the same 
characteristic orbital velocity, but were confined to lie within 
different ranges of inclination to the line of sight. Although the model 
adopted in MD01 provided an excellent fit to the available data, it
was not fully satisfactory. The main reason for this is that because 
there was no way to individually allocate objects to one of the two 
populations, it was therefore only possible to calculate an 
average inclination correction, based on the sample as a whole. 
Consequently, it was therefore inevitable that objects with 
FWHM at the extremes of the observed distribution would not be
properly corrected for inclination, systematically increasing the 
scatter in the resulting black-hole mass estimates.

Consequently, in the analysis of this new larger 72-object AGN sample it
was decided to investigate the possibility of providing an acceptable
description of the sample FWHM distribution using a single population
model, uniquely determined by only three free parameters. This model
makes the assumption that the intrinsic keplerian orbital velocities
of the sample have a gaussian distribution, with mean $\bar{V}_{orb}$
and variance $\sigma_{orb}^{2}$. The third free parameter of the model
 is $\theta_{max}$, which is the maximum allowed angle between the 
line of sight and the disc rotation axis. In the model fit it is
presumed that the angles between the line of sight and the disc 
rotation axis are randomly distributed between zero (pole-on) 
and $\theta_{max}$. In order to relate the intrinsic orbital
velocities to the observed FWHMs we follow the results of the accretion
disc modelling of Rokaki \& Boisson (1999). Rokaki \& Boisson used
an accretion disc model to fit the UV continuum and H$\beta$
line profiles of a sample of 19 Seyfert 1 galaxies (9 of which are
common to our sample) with the fitting process returning estimates of
both central black-hole mass and the disc inclination angle. From
their results for the 9 objects in common to both samples it follows that:
\begin{equation}
FWHM\simeq 2\times sin(\theta) \left[ GM_{bh}/R_{sat}\right]^{0.5}
\end{equation}
\noindent
where $R_{sat}$ is the radius at which the radial disc 
line-flux $F_{l}(r)$ saturates. It was shown 
by Rokaki \& Boisson that the derived values of $R_{sat}$ are in good 
agreement with the broad-line radii $R_{BLR}$ derived from
reverberation mapping observations. In this study we therefore make
the assumption that $R_{BLR}=R_{sat}$, which immediately leads 
to $FWHM=2V_{orb}\times sin(\theta)$. Consequently, in this disc model
the inclination correction factor is simply $f=\frac{1}{2sin(\theta)}$.
\subsection{Modelling the FWHM distribution}
\label{fwhmsec}
\begin{figure}
\centerline{\epsfig{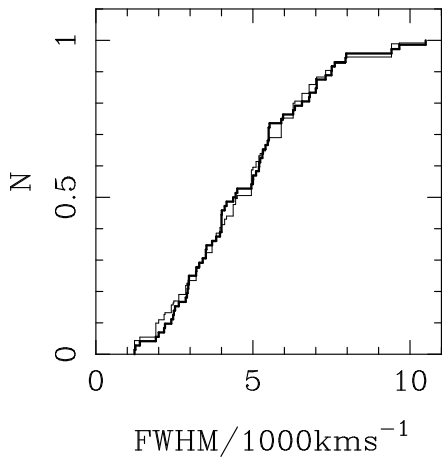}}
\caption{The cumulative FWHM distributions of the AGN sample (thick line)
and the best-fitting disc BLR model (thin line). The two can be seen to be
indistinguishable (KS, $p=0.99$)}
\label{fwhmfig}
\end{figure}
The parameters of the best-fitting model are determined by minimizing
the chi-squared statistic between the model and data FWHM distributions. The 
best-fit parameter values ($\chi^{2}_{min}=2.75, 5$ d.o.f.) 
determined from this process are as follows: $V_{orb}=4375$ kms$^{-1}$, 
$\sigma_{orb}=1400$ kms$^{-1}$ and $\theta_{max}=47^{\circ}$. The
 location of the minimum chi-squared solution, together with the
relevant confidence level contours, is shown in the left-hand panel of
Fig \ref{discmodel}. As a consistency check we have also determined
the best-fit parameters by minimizing the Kolmogorov-Smirnov (KS) statistic
between the model and data cumulative FWHM distributions. The location
of this minimum is also indicated in Fig \ref{discmodel}, and can be
seen to be consistent with the minimum chi-squared solution. The
 quality of the model fit can be seen 
from Fig \ref{fwhmfig}, which shows the cumulative FWHM 
distribution of the full 72-object sample and the best-fitting model. 
The application of the KS test confirms that the two distributions are 
essentially identical ($p=0.99$). 

The second panel of Fig \ref{discmodel} shows the 
inclination factor $f$ predicted by the best-fitting disc model, together with
the predictions of the $\pm{1\sigma}$ models. Also shown in the plot
(horizontal dashed line) is the effective inclination factor resulting
from the assumption of randomly orientated BLR orbits;
$f=\frac{\sqrt{3}}{2}$. The prediction of random orientated BLR orbits
has been included to illustrate that the disc BLR model adopted here
only makes significantly different predictions to the random case for
those objects with observed FWHM$\leq2800$ kms$^{-1}$. For objects with
FWHM$\geq2800$ kms$^{-1}$ ($83\%$ of the 72-object AGN sample) the inclination
 correction factor derived from our disc model differs by less
than a factor of 2 from that resulting from the assumption 
of random velocities. 
\subsection{Testing the geometry of the broad-line region}
\label{geometry}

\begin{table}
\begin{center}
\begin{tabular}{lcl}
\hline
Object&$\sigma$/kms$^{-1}$&Reference\\
\hline
3C  120    &162&NW95\\
MRK 79     &120&F01\\
MRK 590    &169&NW95\\
MRK 817    &140&F01\\
NGC 3227   &128&NW95\\
NGC 3516   &144&A97\\
NGC 4051   &\phantom{0}88 &NW95\\
NGC 4151   &\phantom{0}85 &F01\\
NGC 5548   &180&F01\\
NGC 6814$\star$   &115&NW95\\
\hline
\end{tabular}
\caption{The ten Seyfert galaxies in our sample which have published stellar 
velocity dispersions. All of the objects, with the exception of NGC 6814, 
also have reverberation mapping measurements of $R_{BLR}$. For 
NGC 6814 \,$R_{BLR}$ has been estimated from its continuum luminosity. 
Column three lists the literature sources for the 
velocity dispersions, with the following codes: NW95 (Nelson \& Whittle 
1995), F01 (Ferrarese et al. 2001) and A97 (Arribas et al. 1997).}
\label{seytab}
\end{center}
\end{table}

\begin{figure*}
\centerline{\epsfig{file=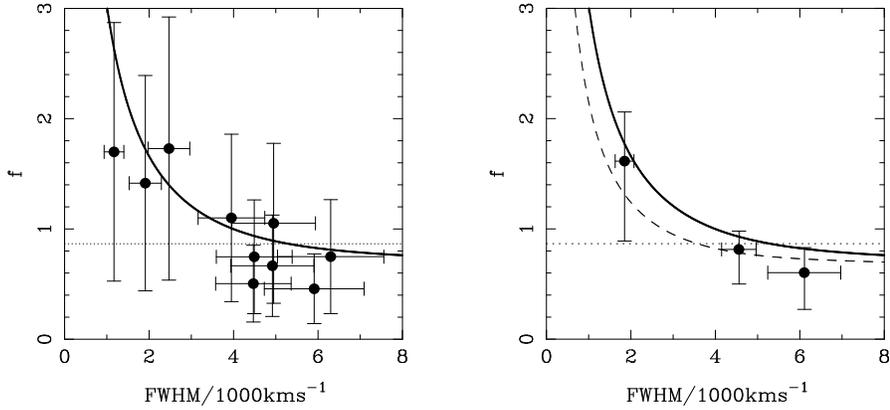,width=12cm,angle=0,clip=}}
\caption{The left-hand panel shows the derived inclination factor
($f$) versus FWHM for the 10 Seyfert galaxies with measured stellar velocity
dispersions (see text for discussion). Also shown is the prediction of 
the best-fitting disc BLR model to the full AGN sample (solid curve)
and the expected value of
$f$ in the case of purely random BLR velocities (dotted 
horizontal line). The error bars have been estimated by assuming a 
scatter of $0.3$ dex in the $M_{bh}-\sigma$ relation and $\pm15\%$
errors in FWHM measurements. The right-hand panel shows the same
information using binning to clarify the trend displayed by the
data. The dashed curve shows the inclination prediction of the
best-fitting disc model to the complete 12$\mu m$ Seyfert 1 sample of 
Rush, Malkan \& Spinoglio (1993), which has a lower mean black-hole 
mass than our full AGN sample (see text for discussion).}
\label{revdisc}
\end{figure*}
Although the inclination corrections introduced by the disc BLR model 
only have a significant effect for a minority of objects in the 
AGN sample, it will be demonstrated in the
next section that they significantly reduce the scatter in 
the $M_{bh}-L_{bulge}$ relation. 

Due to this fact, it is important that
the inclination dependence of the observed FWHM predicted by the 
disc BLR model can be shown to be at least consistent with the
available data. As was discussed in Section \ref{discsec}, although there is
substantial evidence in the literature for an inclination effect in
the observed FWHM of radio-loud AGN, such evidence is not yet available
for radio-quiet AGN. As a result, the justification for 
extending the disc model to radio-quiet objects relies primarily on
the assumption that the central engine of all powerful AGN are 
essentially the same. 

However, included in our 19-object Seyfert galaxy sample are 10
objects for which reliable stellar velocity dispersion measurements are
available in the literature (see Table 2). Because 
stellar velocity dispersions are orientation independent, it
follows that a direct comparison of the black-hole masses predicted by 
the $M_{bh}-\sigma$ relation with those of the orientation 
dependent H$\beta$ method will allow a test of the BLR geometry.

The procedure is as follows. For each of the ten Seyfert galaxies we
first estimate their black-hole mass using the best-fitting $M_{bh}-\sigma$
relation to our nearby inactive galaxy sample (see Section \ref{main}) 
which has the form:
\begin{equation}
\log M_{bh}=4.09(\pm0.58)\log(\sigma) -1.26(\pm1.39)
\end{equation}
\noindent
and can be seen to be consistent with the $M_{bh}-\sigma$ relations 
determined by both Merritt \& Ferrarese ($M_{bh}\propto \sigma^{4.72}$) and 
Gebhardt et al. (2000b)($M_{bh}\propto \sigma^{3.75}$). We then require 
for each object that this stellar velocity dispersion black-hole mass 
estimate be equal to the H$\beta$ virial black-hole mass estimate 
given by Eqn \ref{eqn1}, using the reverberation mapping estimate 
of $R_{BLR}$ and assuming $V=f\times$FWHM. This provides a 
model-independent estimate of the inclination factor $f$ for
each of the 10-objects. In Fig \ref{revdisc} 
we plot the estimated values of $f$ against observed FWHM for 
the ten Seyfert galaxies, along with the predicted curve from 
the disc BLR model, and the effective inclination factor 
resulting from the assumption of randomly orientated BLR velocities. From
Fig \ref{revdisc} there can be seen to be good agreement between the 
derived inclination correction factors and those predicted by 
our adopted disc BLR model, particularly in the right-hand panel where
the data have been binned. Also shown in the right-hand panel of Fig
\ref{revdisc} is the inclination correction curve produced by fitting
to the FWHM distribution of Seyfert galaxies alone. Due to its lower 
characteristic black-hole mass, it can be seen that this curve
represents an improved match to the trend displayed by the binned data. 
We note here that the possible impact of fitting the various AGN 
sub-samples with different characteristic orbital velocities has been 
investigated. However, because the
differences introduced into the black-hole mass estimates are small
($\sim0.1$ dex), this more complicated procedure has negligible effect
on the results presented in the rest of the paper. Consequently it was
decided to adopt the best-fitting model to the full AGN sample throughout.

Although the small
sample size makes it unwise to draw any firm conclusions, we simply
note that the assumption that the observed FWHM of the AGN sample
are inclination dependent is at least consistent with the available
data, and apparently more so than the assumption of 
purely random BLR velocities.
\section{The relationship between bulge luminosity and black-hole mass.}
\label{main}
\begin{figure}
\centerline{\epsfig{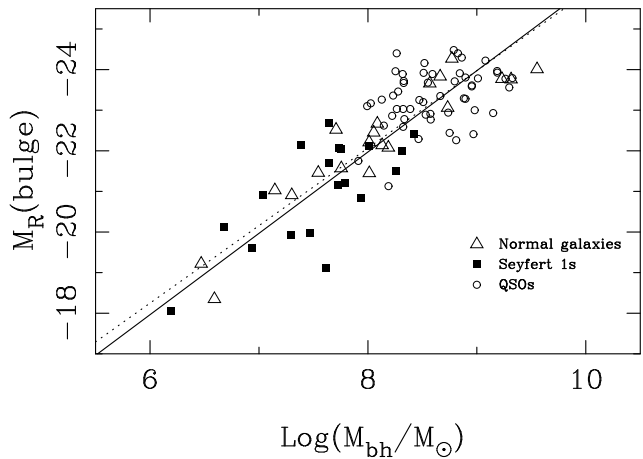}}
\caption{Absolute $R-$band bulge magnitude versus black-hole mass for
the full 90-object sample. The black-hole masses for the 72 AGN are
derived from their H$\beta$ line-widths under the disc-like BLR
model. The black-hole masses of the inactive galaxies (triangles) are
dynamical estimates as compiled by Kormendy \& Gebhardt (2001). Also
shown is the formal best-fit (solid line) and the best-fitting 
linear relation (dotted line).}
\label{mainfig}
\end{figure}

In Fig \ref{mainfig} absolute $R-$band bulge magnitude is plotted
against black-hole mass for the 72 objects in the AGN
sample. Also shown is absolute $R-$band bulge
magnitude plotted against dynamically-estimated black-hole mass for the 20
objects in our nearby inactive galaxy sample. Two aspects of Fig
\ref{mainfig} are worthy of immediate comment. Firstly, as was shown by MD01
and Laor (1998 \& 2001a), it can be seen that bulge luminosity and
black-hole mass are extremely well correlated, with the Spearman rank
correlation test returning $r_{s}=-0.77$ ($7.3\sigma$). Secondly, it
is clear that the AGN and nearby inactive galaxy samples follow the
same $M_{bh}-L_{bulge}$ relation over $>3$ decades in black-hole mass,
and $>2.5$ decades in bulge luminosity. This second fact strongly
supports the conclusions of Dunlop et
al. (2001) and Wisotzki et al. (2001), that the host-galaxies of 
powerful quasars are normal massive ellipticals drawn from the bright
end of the elliptical galaxy luminosity function. Thirdly, there
can be seen to be no systematic offset between the Seyfert 1 and 
quasar samples, reinforcing the findings of MD01 and Laor (2001a)
that, contrary to the results of Wandel (1999), the bulges of 
Seyfert galaxies and QSOs form a continuous
sequence which ranges from $M_{R}$(bulge)$\simeq-18$ to
$M_{R}$(bulge)$\simeq-24.5$. If we adopt an integrated value of 
$M_{R}^{\star}=-22.2$ (Lin et
al. 1998), then this implies that the $M_{bh}-L_{bulge}$
relation holds from $L_{bulge}\simeq 0.01L^{\star}$, all the way up
to objects which constitute some of the most massive ellipticals ever
formed; $L_{bulge}\simeq 10L^{\star}$. 

In order to find the best-fitting $M_{bh}-L_{bulge}$ relation we use 
the same iterative $\chi^{2}$ method used in MD01, which allows for
measurement errors in both coordinates (Press et al. 1992). The
best-fit to the full 90-object sample has the following form:
\begin{equation}
\log(M_{bh}/\Msolar)=-0.50(\pm0.02)M_{R}-2.96(\pm0.48)
\label{bestfit}
\end{equation}
\noindent
and is shown as the solid line in Fig \ref{mainfig}. The scatter around
this best-fitting relation is $\Delta M_{bh}=0.39$ dex. It is worth
noting that this level of scatter 
means that the $M_{bh}-L_{bulge}$ relation is now
worthy of increased interest, given that it amounts to an uncertainty
factor of $<2.5$. Given that the 20 objects in the nearby inactive
galaxy sample have actual dynamical black-hole mass estimates, it is 
obviously of interest to quantitatively test how consistent the 
$M_{bh}-L_{bulge}$ relation for these objects is with the fit to 
the full, AGN dominated, sample. The best-fit to the inactive 
galaxy sub-sample alone, has the following form:
\begin{equation}
\log(M_{bh}/\Msolar)=-0.50(\pm0.05){\rm M}_{\rm R}-2.91(\pm1.23)
\end{equation}
\noindent
which can be seen to be perfectly consistent with the best-fit to the
full sample in terms of both slope and normalization. Indeed, as shown
in Table \ref{fitstab}, the best-fitting relations for the full sample,
quasar sample, Seyfert galaxy sample and the nearby inactive 
galaxy sample are all internally consistent, and display comparable
levels of scatter.  This is a remarkable result 
given that it implies that the combined bulge/black hole formation 
process was essentially the same throughout the full sample, which as well 
as featuring both active and inactive galaxies, includes 
galaxies of both late and early-type morphology.

\begin{figure*}
\centerline{\epsfig{file=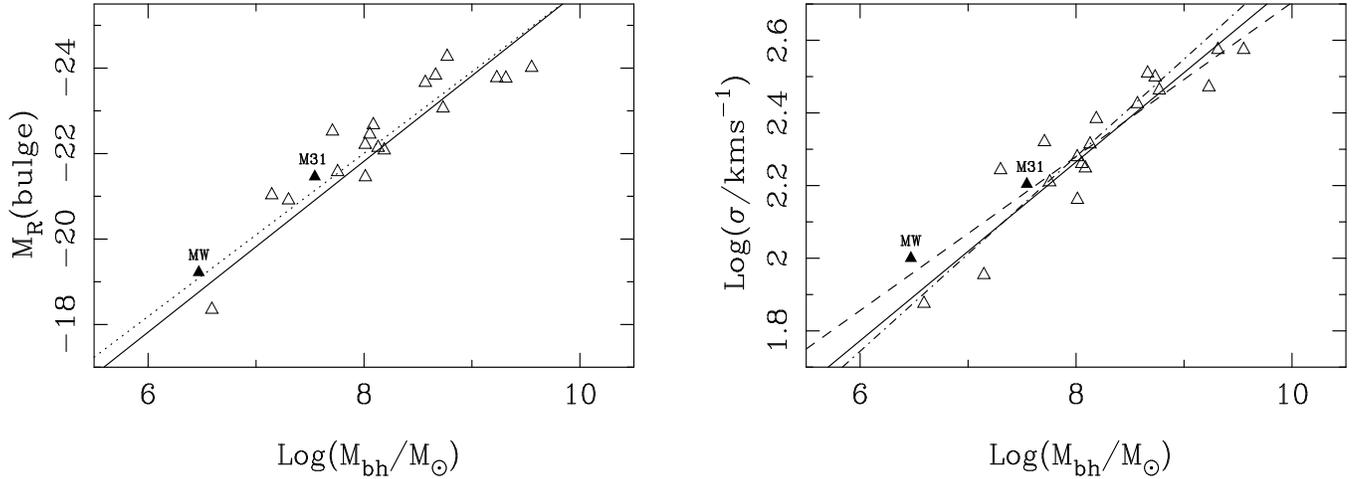,width=18cm,angle=0,clip=}}
\caption{Left-hand panel shows absolute $R-$band bulge magnitude
versus dynamical black-hole mass estimate for our 18-object inactive
galaxy sample. The solid line is the best-fitting relation which has
the form $M_{bh} \propto M_{bulge}^{0.95\pm0.09}$. The dotted
line is the best-fitting linear relation to the full sample, and 
is equivalent to $M_{bh}=0.0012M_{bulge}$. The right-hand panel is the
same with bulge luminosity replaced by stellar velocity
dispersion. The solid line is the best-fit ($M_{bh}\propto \sigma^{4.09}$)
, the dashed line is the 
Merritt \& Ferrarese (2000b) relation ($M_{bh}\propto \sigma^{4.72}$),  and 
the dot-dashed line is the Gebhardt et al. (2000b) relation 
($M_{bh}\propto \sigma^{3.75}$). In both figures the location of the
Milky Way and M31 are indicated for the interest of the reader,
although neither was included in the nearby galaxy sample for
the purposes of the analysis.}
\label{mainfig2}
\end{figure*}

The quality of the fit to the inactive galaxy sample is 
illustrated by the left-hand
panel of Fig \ref{mainfig2}, which shows the $M_{bh}-L_{bulge}$ relation
for the inactive galaxy sample alone. Of particular interest is
the scatter around this best-fit relation, given that it has been widely
reported in the literature (eg. Merritt \& Ferrarese 2000b, Kormendy \&
Gebhardt 2001) that the scatter around the $M_{bh}-L_{bulge}$ relation 
is significantly greater than that around the $M_{bh}-\sigma$
relation. However, in contrast, we find that the scatter 
around the $M_{bh}-L_{bulge}$ relation for our sample of 
nearby inactive galaxies, which excludes non E-type
morphologies, is only $0.33$ dex, in excellent agreement with 
the scatter around the $M_{bh}-\sigma$ 
relation (Merritt \& Ferrarese 2001b).

To test this result further, in the right-hand panel of 
Fig \ref{mainfig2}, we investigate $M_{bh}-\sigma$ relation for 
our nearby inactive galaxy sample. The scatter around the 
best-fit relation ($M_{bh} \propto \sigma^{4.09}$) is 0.30 dex, 
leading us to the conclusion that the intrinsic 
scatter around the $M_{bh}-L_{bulge}$ relation for elliptical galaxies
is comparable to that in the M$_{bh}-\sigma$ relation. This may
largely be due to the fact that when constructing the 
nearby inactive galaxy sample, objects with late-type 
morphologies were deliberately
excluded. This would support the conclusion of MD01 that
successful disc/bulge decomposition of late-type galaxies is a
difficult task, even with high resolution data, and that bulge
luminosities from ground-based images with poor seeing and, in
particular, morphology-based estimates of bulge/total luminosity 
fraction (eg. Simien \& de Vaucouleurs 1986) can often have substantial
systematic errors associated with them. However, we also note that 
the late-type galaxies excluded from our nearby galaxy sample are
not significant outliers with respect to the $M_{bh}-\sigma$ relation.
Consequently, although our results indicate that bulge luminosity can
provide accurate black-hole mass estimates, comparable with the 
$M_{bh}-\sigma$ relation, for elliptical galaxies, 
it is clear that the $M_{bh}-\sigma$ relation will provided more
accurate black-hole mass estimates for samples which are not
restricted to solely ellipticals.
\begin{center}
\begin{table}
\begin{tabular}{lccccc}
\hline
Sample&N&a&b&$\chi^{2}$&$\Delta M_{bh}$\\
\hline
All	&90&$-2.96\pm0.48$&$-0.50\pm0.02$&1.43&0.39\\
Active	&72&$-2.25\pm0.72$&$-0.46\pm0.03$&0.94&0.42\\
QSOs    &53&$-4.26\pm3.15$&$-0.55\pm0.13$&0.87&0.42 \\
Seyfert &19&$-1.96\pm1.33$&$-0.45\pm0.06$&1.21&0.43\\
Inactive&20&$-2.91\pm1.04$&$-0.50\pm0.05$&1.05&0.33\\  
\hline
\end{tabular}
\caption{Details of the best-fitting relations of the form $\log
M_{bh}=a+b M_{R}(bulge)$ to the full sample and to the various 
sub-samples. Column two gives the number of objects included
in each fit. Columns three and four give the best-fitting values of
the parameters $a$ \& $b$, along with their associated
uncertainties. The relatively large uncertainties in the fit to the
QSO sub-sample reflects the small dynamic range in terms of both
$L_{bulge}$ and $M_{bh}$. Column five gives the reduced $\chi^{2}$ for 
each fit and column 6 gives the scatter around each of the 
best-fitting relations in terms of $\Delta \log M_{bh}$.}
\label{fitstab}
\end{table}
\end{center}
\subsection{The linearity of the bulge: black-hole mass relation}
As pointed out in Section \ref{intro}, one of the aims of this paper
was to investigate the linearity of the $M_{bh}-M_{bulge}$ 
relation. In our previous study (MD01) of a sample of 45 AGN we found
that $M_{bh} \propto M_{bulge}^{1.16\pm0.16}$, and therefore
concluded that there was no evidence that the $M_{bh}-M_{bulge}$
relation was non-linear. In contrast, evidence for a non-linear relation was
recently found by Laor (2001a). In his $V-$band study of the black hole to
bulge mass relation in a 40-object sample (15 PG quasars, 16 inactive
galaxies and 9 Seyfert galaxies) Laor found a best-fitting relation of
the form:
\begin{equation}
M_{bh}=M_{bulge}^{1.54\pm0.15}
\end{equation}
\noindent
which is clearly apparently inconsistent with linearity. However, 
in order to determine
the $M_{bh}-M_{bulge}$ relation it is obviously necessary to convert
the measured bulge luminosities into masses, via an adopted
mass-to-light ratio. The form of this mass-to-light ratio affects the derived
slope of the $M_{bh}-M_{bulge}$ relation in the following way. If
the mass-to-light ratio is parameterized as $M/L \propto L^{\alpha}$,
then the resulting slope ($\gamma$) of the $M_{bh}-M_{bulge}$
relation is given by $\gamma=\frac{-2.5 \beta}{1+\alpha}$, where
$\beta$ is the slope of the $M_{bh}-L_{bulge}$ relation (Eqn 5).

As in MD01 we choose to adopt the derived $R-$band mass-to-light ratio
for the Coma cluster from J$\o$rgensen, Franx \& Kj$\ae$rgaard (1996),
which has $\alpha=0.31$. With this mass-to-light ratio the 
best-fitting $M_{bh}-L_{bulge}$ relation (Eqn \ref{bestfit}) 
transforms to a $M_{bh}-M_{bulge}$ relation of the following form:
\begin{equation}
M_{bh} \propto M_{bulge}^{0.95\pm0.05}
\end{equation}
\noindent
It can immediately be seen that 
that there is no indication from our results that the relation 
is inconsistent with a linear scaling between black hole and bulge
mass. 

In order to calculate the bulge mass of the objects in his sample, Laor 
(2001a) adopted a $V-$band mass-to-light ratio of 
$M_{bulge}\propto L_{bulge}^{1.18}$ (Magorrian et al. 1998), which 
is significantly different from our chosen mass-to-light
ratio. Indeed, if Laor (2001a) had adopted the mass-to-light ratio
used here, then his best-fit to the $M_{bh}-M_{bulge}$ relation for 
his full sample would be $M_{bh} \propto M_{bulge}^{1.38\pm0.15}$, 
which is not formally inconsistent with linearity. However,
irrespective of this, our new best-fit to the slope of the 
$M_{bh}-L_{bulge}$ relation ($\beta=-0.50\pm0.02$) of our new sample,
which has a larger dynamic range in $L_{bulge}$ than both the samples
studied in MD01 and Laor (2001a), means that any disagreement about
mass-to-light ratios cannot now alter the conclusion that the
$M_{bh}-M_{bulge}$ relation is consistent with being linear. To
demonstrate this we conclude by noting that even using 
the $M_{bulge}\propto L_{bulge}^{1.18}$ mass-to-light ratio adopted by 
Laor (2001a), our best-fitting $M_{bh}-L_{bulge}$ relation is equivalent to:
\begin{equation}
M_{bh} \propto M_{bulge}^{1.06\pm0.06}
\end{equation}
\noindent
again, completely consistent with a linear scaling. 

\subsection{The normalization of the $\bf{M_{bh}-M_{bulge}}$ relation}
\label{norm}
Having established that the $M_{bh}-M_{bulge}$ relation is consistent 
with being linear, we now assume perfect linearity in order to
establish the normalization of the $M_{bh}-M_{bulge}$ relation. With the 
mass-to-light ratio adopted here, a linear scaling corresponds to 
enforcing a slope of $-0.524$ in the $M_{bh}$ vs. $M_{R}$ relation. 
Under this restriction the best-fitting relation (dotted line in 
Figs \ref{mainfig} \& 5a) has a normalization of
$M_{bh}=0.0012M_{bulge}$, and can clearly be seen to be an 
excellent representation of the data. It is noteworthy that the 
normalization of $M_{bh}=0.0012M_{bulge}$ is identical to that determined by
Merritt \& Ferrarese (2000a) from their velocity dispersion study of
the 32 inactive galaxies in the Magorrian et al. sample. 

\begin{figure}
\centerline{\epsfig{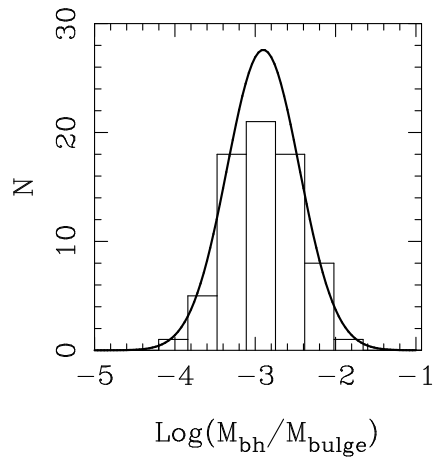}}
\caption{Histogram of the ratio of black-hole mass to bulge mass for
the 72-object AGN sample. Over-plotted for comparison is a gaussian
with $\langle \log (M_{bh}/M_{bulge}) \rangle =-2.90$ and standard deviation 
0.45 (see text for discussion)}
\label{ratio}
\end{figure}

The closeness of the
 agreement between the $M_{bh}/M_{bulge}$ ratios determined here with
 those determined by Merritt \& Ferrarese is highlighted by Fig
 \ref{ratio}, which shows a histogram of the $M_{bh}/M_{bulge}$
 distribution for our 72-object AGN sample. The AGN $M_{bh}/M_{bulge}$ 
distribution has $\langle \log (M_{bh}/M_{bulge})\rangle =-2.87
 \pm{0.06}$ with a standard deviation of $\sigma=0.47$. This is in 
remarkably good agreement with the Merritt \& Ferrarese results, which were 
$\langle \log (M_{bh}/M_{bulge}) \rangle =-2.90$ and $\sigma=0.45$. 

The close agreement between the distribution of $M_{bh}/M_{bulge}$ 
found here for the 72-object AGN sample and that found by Merritt \& Ferrarese 
using velocity dispersions of nearby inactive galaxies, can also be taken 
as further supporting evidence that the adoption of a disc model for 
the BLR geometry is valid. Finally, we note that the normalization of 
$M_{bh}=0.0012M_{bulge}$ agrees very well with the predictions of 
recent models of coupled bulge/black hole formation at high redshift 
(Archibald et al. 2001).

\section{The bolometric luminosity -- black-hole mass correlation}
\label{leddsec}
\begin{figure}
\centerline{\epsfig{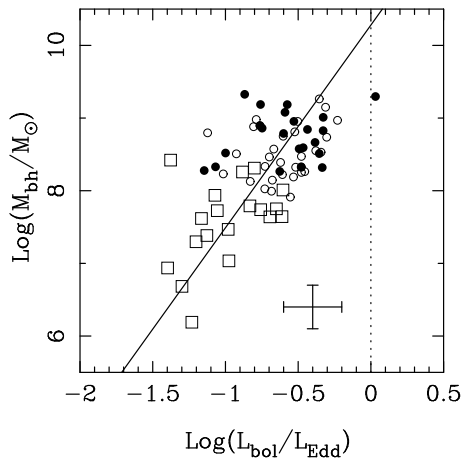}}
\caption{Black-hole mass plotted against bolometric luminosity as a
fraction of the Eddington limit for the 72-object AGN sample. Radio-loud 
quasars are shown as filled circles, radio-quiet quasars are shown as open 
circles and the Seyfert galaxies are shown as open squares. The
best-fitting relation is shown as the solid line, with the location of
the Eddington limit shown by the vertical dotted line. Also shown is a 
representative error bar.}
\label{ledd}
\end{figure}
In this section we use the optical continuum information available for
the AGN sample to investigate how bolometric luminosity as a fraction
of the Eddington limit varies with black-hole mass. In order to
estimate the bolometric luminosity of each AGN we follow Wandel,
Peterson \& Malkan (1998) and adopt $L_{bol}\simeq10
\lambda L_{5100}$, where $L_{5100}$ is the monochromatic luminosity at
5100 \AA. Although this calibration was also found to be appropriate 
for the composite quasar spectrum of Laor et al. (1997), it is
obviously only a rough estimate (Wandel, Peterson \& Malkan 1998) 
and consequently we allow for a factor of 4 uncertainty when 
deriving the best-fitting relation.

In Fig \ref{ledd} we plot black-hole mass against 
$L_{bol}/L_{Edd}$ for the full 72-object AGN sample. It can be
seen that the two quantities are well correlated, with a Spearman rank
coefficient of $r_{s}=0.50, 4.2\sigma$. The best-fitting relation
(solid line) has the following form:
\begin{equation}
\log M_{bh}=2.79(\pm0.37)\log(L_{bol}/L_{edd}) -10.29(\pm0.28)
\end{equation}
\noindent
and shows clearly that the data are inconsistent with all AGN having a
constant Eddington ratio, a conclusion which was also arrived at by 
Kaspi et al. (2000). As in Fig \ref{mainfig}, it is apparent that the
Seyfert galaxies and quasars form a continuous sequence, and that in terms of 
Eddington ratio, there is no evidence for any systematic
offset between the radio-loud and radio-quiet quasars. However, it is
noteworthy that there is no evidence for a correlation within 
the QSO sub-sample alone. Indeed, given that the observed correlation is
entirely dependent upon the heterogeneous Seyfert galaxy sample, it
should perhaps be viewed with some caution. For example, it is
inevitable that objects which should occupy the top-left corner of Fig
\ref{ledd} will be missing from our sample, since their large bulge
luminosity and weak nuclear emission will have prevented them from
being classified as powerful AGN. However, it is not immediately
obvious why there should be no members of the Seyfert galaxy sample
which are radiating at much closer to their Eddington limit. 
Although it is possible that the bottom-right of Fig \ref{ledd} could 
be populated by so-called narrow-line Seyfert galaxies, which are 
thought to have intrinsically small black-hole masses and to radiate 
at a large fraction of their Eddington limit (eg. Mathur 2000), 
our present study does not support this. Four of the objects in the 
Seyfert galaxy sample have FWHM$\le2200$ kms$^{-1}$ and yet do not 
occupy this region of the diagram. The implications of the
$M_{bh}-L_{bol}/L_{Edd}$ relation, together with the $M_{bh}-M_{bulge}$
relation determined in the previous section, will be explored in a
forthcoming paper (McLure, Percival \& Dunlop, in prep).

\section{black-hole mass and the radio-loudness dichotomy}
\label{dichotomy}
In this final section we investigate the relationship between black-hole mass
and radio luminosity for the objects in the AGN sample, to explore the role 
played by black-hole mass in the quasar radio-loudness dichotomy. To ensure 
that any connection between black-hole mass and radio power can be 
investigated properly it is essential that the radio-quiet and radio-loud 
sub-samples should be indistinguishable in terms of both redshift and 
optical nuclear luminosity. From the total of 53 quasars in
 the AGN sample it has been possible to construct two sub-samples of 
radio-loud and radio-quiet quasars, numbering 23 and 26 objects 
respectively, which are well matched in term of redshift and 
optical luminosity ($\lambda L_{5100}$). The matching of the two 
sub-samples is confirmed by the application of the two-dimensional KS test 
(Peacock 1983), which returns a probability of $p=0.25$ that the two 
distributions are indistinguishable in the $\lambda L_{5100}-z$ plane. Given 
that the quasars as a whole are a heterogeneous sample, this 
$\lambda L_{5100}-z$ matching is the best that can be done to 
ensure that any differences in the sub-sample black-hole mass 
distributions should be linked to the large differences in radio power.

\begin{figure}
\centerline{\epsfig{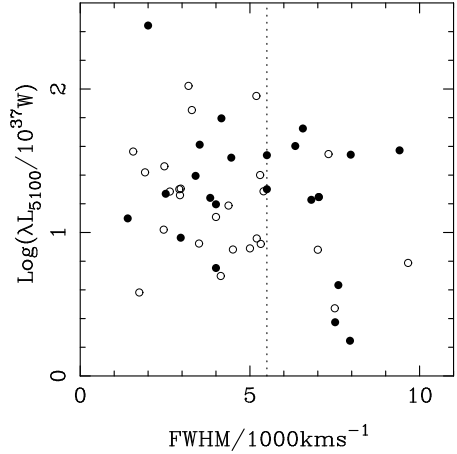}}
\label{matched}
\caption{The distribution of the radio-loud and radio-quiet quasar sub-samples 
in the $\lambda L_{5100}-$FWHM plane. It should be noted that
there is a significant difference in the FWHM distributions of the two
samples (KS, $p=0.05$), with 11/23 radio-loud quasars having 
FWHM$>5500$ kms$^{-1}$ while only 4/26 radio-quiet quasars 
have FWHM$>5500$ kms$^{-1}$.}
\label{matchedfig}
\end{figure}

The mean black-hole mass of the radio-quiet quasar sample is 
$\langle \log (M_{bh}/\Msolar) \rangle =8.61\pm0.07$, with a median value 
of $\log (M_{bh}/\Msolar)=8.56$. The black-hole 
masses of the radio-loud quasar sub-sample are somewhat larger on average, 
with a mean of $\langle \log (M_{bh}/\Msolar) \rangle =8.76\pm0.07$,
and a median of $\log (M_{bh}/\Msolar)=8.83$. The suggestion from 
these results is that there 
does appear to be a tendency for the radio-loud quasars to have larger 
black-hole masses than their radio-quiet counterparts 
(the median figures differ by nearly a factor of 2), although the 
overlap between the two sub-samples is 
sufficient to ensure that the two black-hole mass distributions are not 
statistically distinguishable (KS, $p=0.21$). Given that the radio-loud and 
radio-quiet sub-samples were deliberately chosen to be matched in terms of 
their nuclear optical luminosity, it is clear from Eqn \ref{eqn2} that 
the difference in average black-hole mass between the two sub-samples must 
be due to differences in observed H$\beta$ FWHM. This does appear to be the 
case, and can be seen in Fig \ref{matchedfig}, which shows a plot of optical 
luminosity against observed FWHM for the two sub-samples. As is 
highlighted in Fig \ref{matchedfig} there is a separation 
around FWHM=5500 kms$^{-1}$, with 11/23 RLQs having 
FWHM $>5500$ kms$^{-1}$, while only 4/26 RQQs have
FWHM$>5500$ kms$^{-1}$. This difference is confirmed as being 
significant, with the KS test returning a probability of $p=0.05$ that 
the two FWHM distributions are drawn from the same parent distribution. The
difference in the FWHM distributions between radio-loud and radio-quiet
quasars has been previously reported at higher significance
($p=0.001$) from the larger-scale spectrophotometric studies of Corbin
(1997) and Boroson \& Green (1992).

The suggestion that the black holes of radio-loud quasars are on average 
more massive than their radio-quiet counterparts is consistent with the
results of our recent host-galaxy study (Dunlop et al. 2001, McLure et
al. 1999), our previous H$\beta$ line-width study (MD01) and the 
recent line-width study of the LBQS by Laor (2001b). In this study, Laor 
calculated virial black-hole mass estimates for the 87 LBQS objects with 
$z<0.5$, finding an apparent separation between the radio-loud and 
radio-quiet quasars at a black-hole mass of
$\simeq10^{9}\Msolar$. 
However, it is important to remember that the radio-loud objects in 
both of these 
studies are predominantly optically selected. The possible weakness of 
using optically selected quasar samples to study the link between black-hole 
mass and radio power has recently been highlighted by Lacy et al. (2001). In 
their line-width study of a virtually complete sub-sample of the FBQS,
 Lacy et al. find that the apparent gap in the radio: black-hole mass plane 
is filled in by radio intermediate quasars, and that the quasar population 
as a whole follows a relation of the 
form: $P_{5GHz} \propto M_{bh}^{1.9}(L/L_{edd})^{1.0}$. 

Although the black-hole masses and 5GHz radio luminosities of the 
radio-loud quasars analysed here are consistent with the 
results of Lacy et al., the large scatter around the 
Lacy et al. relation (1.1 dex) makes this somewhat 
inevitable. In fact, we argue elsewhere (Dunlop et al. (2001), McLure
\& Dunlop, in prep) that the data are more consistent with the
existence of an upper and lower envelope to the radio power that can
be produced by a black hole of a given mass. Consequently, we argue
that in order to produce a truly radio-loud AGN ($P_{5GHz}\simeq
10^{25}$ WHz$^{-1}$sr$^{-1}$), the central black-hole mass must be
$\ge10^{8.5}$\Msolar\, and lie on the upper $M_{bh}-P_{5GHz}$ 
envelope (see Dunlop et al. (2001) for a full discussion). Finally, 
we note that this picture of the radio-loudness dichotomy is in good 
agreement with the predictions of the merger model proposed 
by Wilson \& Colbert (1995), in which the radio-loudness of an AGN 
is dependent on both the mass and angular momentum of the central black hole.

\section{Conclusions}
\label{conclusions}
New virial black-hole mass estimates have been presented for a sample of 
72 AGN using the FWHM of the broad H$\beta$ emission line, corrected for 
inclination under the assumption of a disc-like broad-line region. The
form of the $M_{bh}-M_{bulge}$ relation is investigated using reliable 
bulge luminosity measurements for both the AGN sample and a carefully 
selected sample of nearby inactive galaxies with dynamical black-hole 
mass measurements. Using the 
inactive nearby galaxy sample a rigorous comparison was made of the scatter 
around the $M_{bh}-L_{bulge}$ and $M_{bh}-\sigma$ relations. 
The main conclusions of this study can be summarized as follows:
\begin{itemize}
\item{For a 10-object subset of the Seyfert galaxy sample it is found that 
the inclination predictions of a disc model BLR are more consistent 
with the available data than the assumption of purely random 
broad-line velocities.}
\item{A strong correlation ($7.3\sigma$) is found between bulge
luminosity and black-hole mass estimated via the disc BLR model.}
\item{The best-fitting  $M_{bh}-L_{bulge}$ relation to the combined 
sample of 72 AGN and 18 nearby inactive elliptical galaxies is found to be 
consistent with a linear scaling between black hole and bulge mass
($M_{bh}\propto M_{bulge}^{0.95\pm0.05}$), and to have much lower
scatter than previously reported ($\Delta \log M_{bh}=0.39$ dex).}
\item{The best-fitting normalization of the $M_{bh}-M_{bulge}$ relation is 
found to be $M_{bh}=0.0012M_{bulge}$, in excellent agreement with 
recent stellar velocity dispersion studies.}
\item{In contrast to previous reports it is found that the
scatter around the $M_{bh}-L_{bulge}$ and $M_{bh}-\sigma$ relations
for the nearby inactive elliptical galaxy sample are comparable, at
only $\sim 0.3$ dex. }
\item{Using samples matched in terms of optical luminosity, the median
black-hole mass of radio-loud quasars is found to be larger by 0.27
dex than that of their radio-quiet counterparts. This difference is found to 
be due to a small but significant ($p=0.05$) difference in their 
respective FWHM distributions.}
\end{itemize}
\section{acknowledgments}
The United Kingdom Infrared Telescope is operated by
the Joint Astronomy Centre on behalf of the U.K. Particle Physics and
Astronomy Research Council. Based on observations with
the NASA/ESA Hubble Space Telescope, obtained
at the Space Telescope Science Institute, which is operated by the
Association of Universities for Research in Astronomy, Inc. under NASA
contract No. NAS5-26555.
This research has made use of the NASA/IPAC Extragalactic Database (NED)
which is operated by the Jet Propulsion Laboratory, California Institute
of Technology, under contract with the National Aeronautics and Space
Administration. RJM acknowledges the award of a PPARC postdoctoral
fellowship. JSD acknowledges the enhanced research time afforded 
by the award of a PPARC senior research fellowship.


\begin{thebibliography}{46}
\bibitem{1} Archibald E.N., Dunlop J.S., Jimenez R., 
Friaca A.C.S., McLure R.J., Hughes D.H., 2001, MNRAS, 
submitted, astro-ph/0108122
\bibitem{2} Arribas S., Mediavilla E., Gar\'{c}ia-Lorenzo B., Del 
Burgo C., 1997, ApJ, 490, 227
\bibitem{3} Baggett W.E., Baggett S.M., Anderson K.S.J., 1998, AJ, 116, 1626
\bibitem{4} Boroson T.A., Green R.F., 1992, ApJS, 80,109
\bibitem{5} Brotherton M.S., ApJS, 1996, 102, 1 
\bibitem{6} Corbin M.R., 1997, ApJS, 113, 245
\bibitem{7} Dunlop J.S., McLure R.J., Kukula M.J., Baum S.A., 
O'Dea C.P., Hughes D.H., 2001, MNRAS, submitted, astro-ph/0108397
\bibitem{8} Faber et al., 1997, AJ, 114, 1771
\bibitem{9} Forster K., Green P.J, Aldcroft T.L., Vestergaard M., Foltz C.B.,
Hewett P.C., 2001, ApJS, 134, 35
\bibitem{10} Ferrarese L., Pogge R.W., Peterson B.M., Merritt D., 
Wandel A., Joseph C.L., 2001, ApJ, 555, L79
\bibitem{11} Franceschini A., Vercellone S., Fabian A.C., 1998, MNRAS, 297, 817
\bibitem{12} Fukugita M., Shimasaku K., Ichikawa T., 1995, PASP, 107, 945
\bibitem{13} Gebhardt K., et al., 2000b, ApJ, 539, L13
\bibitem{14} Gebhardt K., et al., 2000a, ApJ, 543, L5
\bibitem{15} Hooper E.J., Impey C.D., Foltz C.B., 1997, ApJ, 480, L95
\bibitem{16} J$\o$rgensen I., Franx M., Kj$\ae$rgaard P., 1996, MNRAS, 280, 167
\bibitem{17} Kaspi S., Smith P.S., Netzer H., Maoz D., Jannuzi B.T., Giveon U.,
2000, ApJ, 533, 631
\bibitem{18} Kauffmann G., Haehnelt M., 2000, MNRAS, 311, 576
\bibitem{19} Kormendy J., Gebhardt K., 2001, astro-ph/0105230
\bibitem{20} Krolik J.H., 2001, ApJ, 551, 72 
\bibitem{21} Lacy M., Laurent-Muehleisen S.A., Ridgway S.E., 
Becker R.H., White R.L., 2001, ApJ, 551, L17
\bibitem{22} Laor A., 1998, ApJ, 505, L83
\bibitem{23} Laor A., 2001b, ApJ, 543, L111
\bibitem{24} Laor A., 2001a, ApJ, 553, 677
\bibitem{25} Lin H., Kirshner P.P., Schectman S.A., Landy S.D., 
Oemler A., Tucker
\bibitem{26} McLure R.J., Dunlop J.S., Kukula M.J., 
Baum S.A., O'Dea C.P., Hughes D.H., 1999, MNRAS, 308, 377
\bibitem{27} McLure R.J., Dunlop J.S., 2001, MNRAS, in press, astro-ph/0009406
\bibitem{28} Magorrian J., et al., 1998, AJ, 115, 2285
\bibitem{29} Mathur S., 2000, MNRAS, 314, L17
\bibitem{30} Merritt D., Ferrarese L., 2001a, MNRAS, 320, L30
\bibitem{31} Merritt D., Ferrarese L., 2001b, ApJ, 547, 140
\bibitem{32} Nelson C.H., Whittle M., 1995, ApJS, 99, 67
\bibitem{33} Peacock J.A., 1983, MNRAS, 202, 615
\bibitem{34} Percival W.J., Miller L., McLure R.J., Dunlop J.S., 
2001, MNRAS, 322, 843
\bibitem{35} Peterson B.M., Wandel A., ApJ, 2000, 540, L13
\bibitem{36} Press W.H., Teukolsky S.A., Vetterling W.T., Flannery B.P., 1992,
Numerical Recipes, Cambridge University Press
\bibitem{37} Rokaki E., Boisson C., 1999, MNRAS, 307, 41
\bibitem{38} Rush B., Malkan M., Spinoglio L., 1993, ApJS, 89, 1
\bibitem{39} Simien F., de Vaucouleurs G., 1986, ApJ, 302, 564
\bibitem{40} Urry M., Padovani P., 1995, PASP, 107, 803
\bibitem{41} V\'{e}ron-Cetty, M.P.; V\'{e}ron, P., 2001, A\&A, 374, 92 
\bibitem{42} Vestergaard M., Wilkes B.J., Barthel P.D., 2000, ApJ, L103 
\bibitem{43} Virani S.N., De Robertis M., VanDalfsen M.L., 2000,AJ, 120, 1739
\bibitem{44} Wandel A., 1999, ApJ, 519, L39
\bibitem{45} Wandel A., Peterson B.M., Malkan M.A., 1999, ApJ, 526,579
\bibitem{46} Wills B.J., Browne I.W.A., 1986, ApJ, 302, 56 
\bibitem{47} Wilson A.S., Colbert E.J.M., 1995, ApJ, 438, 62
\bibitem{48} Wisotzki L., Kuhlbrodt B., Jahnke K., astro-ph/0103112   
\end{thebibliography}
\end{document}